\documentclass[useAMS,usenatbib,usegraphicx]{mn2e}

\usepackage{upgreek}
\usepackage{amsmath}
\usepackage{txfonts}
\usepackage{booktabs}
\usepackage[usenames, dvipsnames]{color}
\usepackage{array}
\usepackage{graphicx}
\usepackage{epstopdf}
\usepackage{bm}
\usepackage{cmap}
\usepackage{tabularx}
\usepackage{multicol}
\usepackage{natbib}
\usepackage{float}
\usepackage{multirow}
\usepackage{ulem}

\begin{document}

\title[Neutron stars: neutrino luminosity \& heat capacity]{Analytic approximations of neutrino luminosities and heat
capacities of neutron stars with nucleon cores}

\author[D.D. Ofengeim et al.]{
D. D. Ofengeim$^{1,3}$\thanks{E-mail: ddofengeim@gmail.com},
M. Fortin$^2$,
P. Haensel$^{2}$,
D. G. Yakovlev$^{3}$,
J. L. Zdunik$^2$\\
$^{1}$ St~Petersburg Academic University, Khlopina 8/3,
St~Petersburg 194021, Russia\\
$^{2}$ N. Copernicus Astronomical Center, Bartycka 18, 00-716 Warsaw, Poland\\
$^{3}$ Ioffe Institute, Politekhnicheskaya 26, St~Petersburg 194021,
Russia}

\date{Accepted . Received ; in original form}
\pagerange{\pageref{firstpage}--\pageref{lastpage}} \pubyear{2016}

\maketitle

\begin{abstract}
We derive analytic approximations of neutrino luminosities and heat
capacities of neutron stars with nucleon cores valid for a wide
class of equations of state of dense nucleon matter. The neutrino
luminosities are approximated for the three cases in which they are
produced by (i) direct Urca or (ii) modified Urca processes in
non-superfluid matter, or (iii) neutrino-pair bremsstrahlung in
neutron-neutron collisions (when other neutrino reactions are
suppressed by strong proton superfluidity). The heat capacity is
approximated for the two cases of (i) non-superfluid cores and (ii)
the cores with strong proton superfluidity. The results can greatly
simplify numerical simulations of cooling neutron stars with
isothermal interiors at the neutrino and photon cooling stages as
well as simulations of quasi-stationary internal thermal states of
neutron stars in X-ray transients. For illustration, a
model-independent analysis of thermal states of the latter sources
is outlined.
\end{abstract}


\section{Introduction}

It is well known that observations of thermal radiation from cooling
(isolated) neutron stars and from accreting neutron stars in
quiescent states of X-ray transients (XRTs) can be used to infer
(constrain) parameters of neutron stars and explore properties of
superdense matter in their interiors (e.g.,
\citealt{YakHaens2003,YLH2003,YakPeth2004,PLPS2009,Pot2015} and
references therein). This can be done by modeling thermal states and
evolution of isolated and transiently accreting neutron stars and by
comparing theoretical models with observations.

In this way one can test various model equations of state (EOSs) of
superdense matter in neutron star cores --- whether these EOSs are
consistent with the data or not. Direct numerical modeling for the
different EOSs and masses of neutron stars is often time consuming
and requires complicated computer codes. It is the aim of the
present paper to simplify the task by obtaining analytic
approximations for basic integral properties of neutron stars which
determine their thermal evolution.

Specifically, we consider thermally relaxed neutron stars which are
isothermal inside (e.g. \citealt{GS1980}). Taking into account the
effects of General Relativity, they should possess spacially
constant redshifted internal temperature
\begin{equation}
\label{eq:Tgrav}
\widetilde{T}=T \,\exp \Phi,
\end{equation}
where $T$ is a local temperature of the matter and $\Phi$ is the
metric function which determines gravitational redshift. A
noticeable temperature gradient persists only in a thin outer (heat
blanketing) envelope with a rather poor thermal conductivity; its
thickness does not exceed a few hundred meters (e.g.,
\citealt{GPE1983,PCY1997}). We will study redshifted integrated
neutrino luminosities $L_\nu^\infty$ and heat capacities $C$ of such
stars which are the main ingredients for simulating their thermal
structure and evolution. These quantities are mainly determined by
bulky and massive neutron star cores and depend on $\widetilde{T}$.
They depend also on the EOS of neutron star matter and on the
stellar mass $M$. The neutrino luminosity $L_\nu^\infty$ is the sum
of the luminosities provided by different neutrino emission mechanisms.
Both quantities,  $L_\nu^\infty$ and $C$, can be strongly affected
by superfluidity in the neutron star cores.

We study a wide class of EOSs of neutron star cores composed of
neutrons, protons, electrons and muons. These EOSs may open or
forbid the powerful direct Urca processes of neutrino emission in
neutron star cores \citep{LPPH1991}. We consider either
non-superfluid (normal) cores or the cores with strong proton
superfluidity which suppresses all neutrino processes involving
protons as well as the proton heat capacity in the core (e.g.
\citealt{Yak2001}). We will calculate $L_\nu^\infty$ and $C\approx
C_{\rm core}$ and approximate them by analytic expressions which are
universal for all chosen EOSs. This universality greatly simplifies
theoretical analysis of observational data on cooling isolated and
transiently accreting neutron stars. For illustration, we give a
sketch of such an analysis for quasi-stationary neutron stars in
XRTs. In our analysis we neglect possible effects of magnetic
fields.

\section{Cooling properties of neutron stars}
\label{sec:2}

Let us outline the main elements of the cooling theory of neutron
stars (e.g., \citealt{YakPeth2004}). During the first $\sim 10^5$~yr
after their birth neutron stars cool mostly via neutrino emission
from their interiors. During an initial cooling period $\sim
10-100$~yr the internal regions of the star become isothermal. Since
then neutrinos are mainly generated in the stellar core. The basic
neutrino emission mechanisms in the core are the \textit{direct
Urca} (DU) and \textit{modified Urca} (MU) processes as well as
\textit{neutrino-pair bremsstrahlung in baryon collisions}. Neutrino
emissivities of these processes strongly depend on the properties of
the matter --- on composition and superfluidity of baryons (e.g.
\citealt{Yak2001}).

In the present paper we assume that the neutron star core consists
of neutrons with some admixtures of protons, electrons and muons
($npe\mu$-matter) in beta equilibrium. The neutrino emissivities of
relevant processes are reviewed, for instance, by \citet{Yak2001}.
We will study the case of fully non-superfluid (non-SF) matter and the
case of strongly superfluid (SF) protons with other particles being in
normal states. The first case refers to the standard neutrino
emission level and describes the so called standard neutrino
candles. The main contribution to the luminosity of the
neutron star core comes from the DU or MU processes. For the MU process,
we have
\begin{equation}
\label{eq:Qmu}
Q_{\rm MU} = Q_{\rm MU\,0} \left( \frac{n_{\rm p}}{n_0} \right)^{1/3} T_9^8
\Omega\left( n_{\rm n}, n_{\rm p}, n_{\rm e}, n_{\mu} \right),
\end{equation}
where $n_{\rm n}$, $n_{\rm p}$, $n_{\rm e}$ and $n_{\mu}$ are the
number densities of neutrons, protons, electrons and muons,
respectively; $n_0 = 0.16$~fm$^{-3}$ is the standard number density
of nucleons in saturated nuclear matter, $T_9$ is a local
temperature expressed in $10^9$~K and $\Omega\sim 1$ is a
dimensionless factor to account for different branches of the
process (see, e.g., \citealt{Yak2001,KYH2016}). In this work we need
only the main dependence $Q_{\rm MU} \propto n_{\rm p}^{1/3}$. The
factor $Q_{\rm MU\,0} \approx 1.75\times
10^{21}$~erg~cm$^{-3}$~s$^{-1}$ is calculated under the assumptions
described by \citet{Yak2001}, with an effective masses of protons
and neutrons $m_{\rm p}^* = 0.7 m_{\rm p}$ and $m_{\rm n}^* = 0.7
m_{\rm n}$, respectively.

For the DU process, we have (e.g., \citealt{Yak2001})
\begin{equation}
\label{eq:Qdu}
Q_{\rm DU} = Q_{\rm DU\,0} \left( \frac{n_{\rm e}}{n_0} \right)^{1/3}
T_9^6 \left( \Theta_{\rm npe} + \Theta_{\rm np\mu} \right),
\end{equation}
where $Q_{\rm DU\,0} \approx 1.96\times 10^{27}$~erg~cm$^{-3}$~s$^{-1}$,
while the factors $\Theta_{\rm npe}$ and $\Theta_{\rm np\mu}$
are equal to $1$ (open the electron and muon processes,
respectively) if Fermi momenta of reacting particles satisfy the
triangle condition; otherwise, these factors are zero. Because of
the triangle conditions, the electron and muon DU processes have
thresholds and can operate only in the central regions of massive
neutron stars. For some EOSs, they do not operate at all.

The SF case corresponds to the most slowly cooling
neutron stars (e.g., \citealt{Of2015-NS}), where all
neutrino reactions involving protons are strongly suppressed by
proton superfluidity and the most efficient neutrino emission is
provided by the nn-bremsstrahlung. Then (e.g. \citealt{Yak2001})
\begin{equation}
\label{eq:Qnn}
Q_{\rm nn} = Q_{\rm nn \, 0} \left( \frac{n_{\rm n}}{n_0} \right)^{1/3} T_9^8,
\end{equation}
where $Q_{\rm nn\,0} \approx 1.77\times 10^{19}$~erg~cm$^{-3}$~s$^{-1}$.

\begin{figure*}
\begin{minipage}[t]{\columnwidth}
\includegraphics[width=\textwidth]{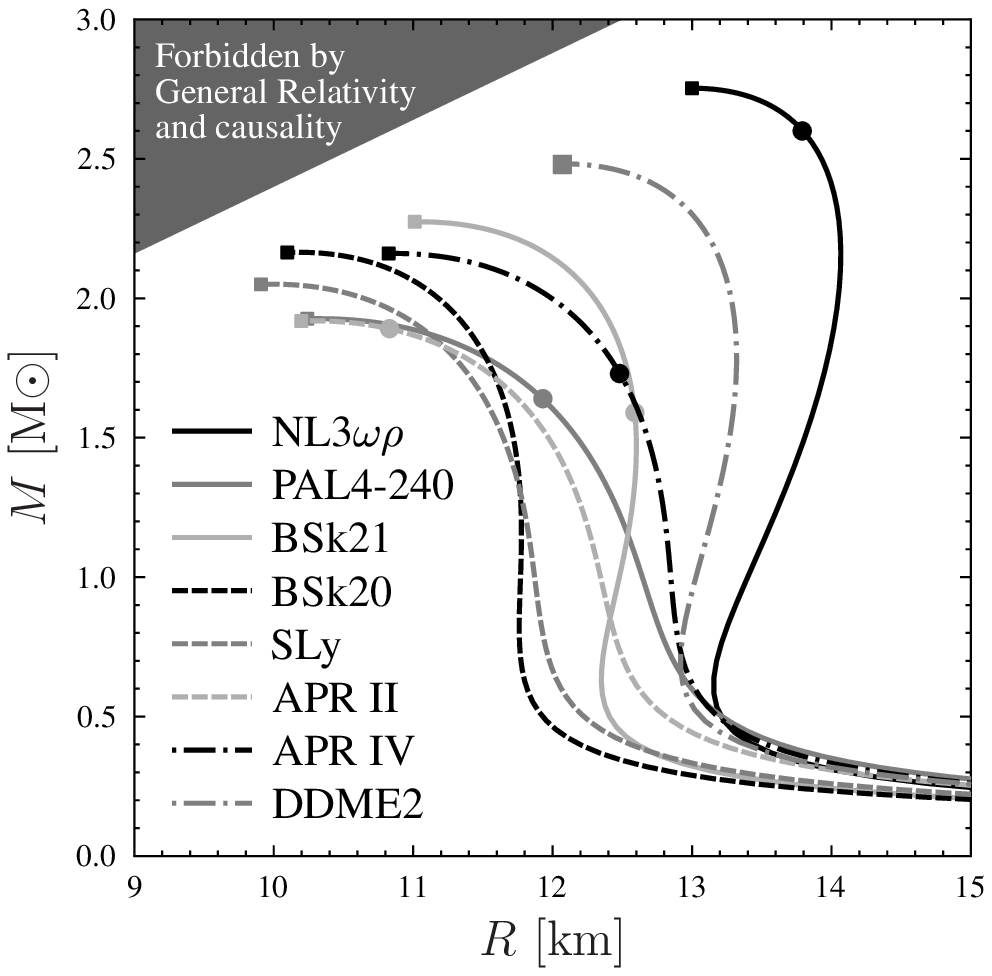}
\caption{\label{fig:M-R} $M-R$ relations for the selected EOSs.
Filled squares mark maximum masses of stable neutron stars with
respective EOSs; filled circles mark minimum masses of stars where DU process is open.}
\end{minipage}
\hfill
\begin{minipage}[t]{\columnwidth}
\includegraphics[width=\textwidth]{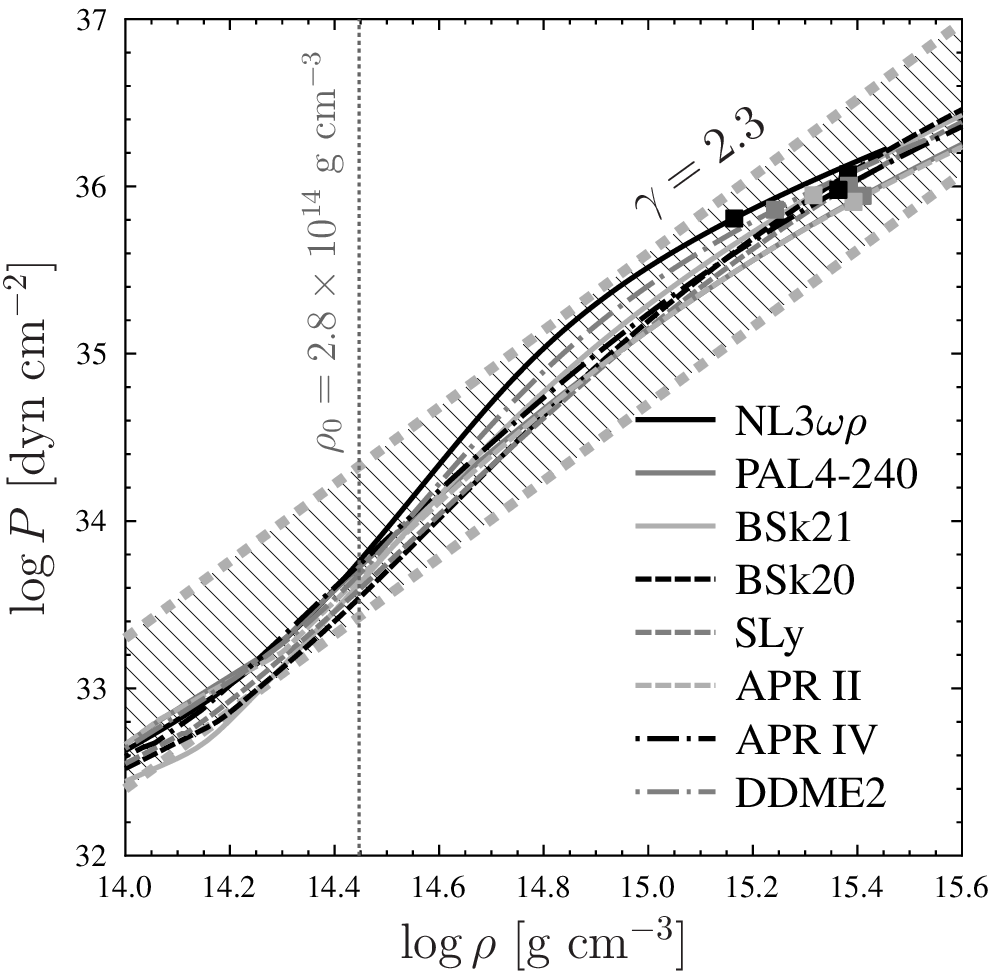}
\caption{\label{fig:P-rho} $P-\rho$ relations for the selected EOSs. Squares
mark the maximum central densities of stable neutron stars with respective EOSs.
The thick shaded strip shows a family of polytropic EOSs with $P \propto \rho^{\gamma}$, where $\gamma=2.3$ is the overall mean value. See text for details.}
\end{minipage}
\end{figure*}

Another quantity important for cooling neutron stars is the
specific heat of the neutron core ${c_{\rm core}}$. Since
neutron stars are mainly composed of strongly degenerate particles,
the heat capacities at constant volume and pressure are almost
identical and we do not discriminate between them. In the non-SF
case there are four contributions to the specific heat in a nucleon
neutron star core,
\begin{equation}
\label{eq:Cv-contrib}
c_{\text{core}} = c_{\rm n} + c_{\rm p} + c_{\rm e} + c_{\mu}.
\end{equation}

For $a =\rm n,\, p,\, e,\, \mu$ one has
\begin{equation}
\label{eq:Cva}
c_{a} = \frac{k_{\rm B}^2}{3\hbar^3} T m_a^* p_{\mathrm{\rm F}a},
\end{equation}
where $k_{\rm B}$ is the Boltzmann constant while $m_a^*$ and
$p_{\mathrm{F}a}$ are, respectively, the effective mass and the
Fermi momentum of particles $a$.
Note that the main contributions
to $c_{\rm core}$ comes from $c_{\rm n}$ and $c_{\rm p}$
(e.g., \citealt{Page1993}). Taking $m_{\rm n}^* =
0.7 \,m_{\rm n}$ and $m_{\rm p}^* = 0.7\, m_{\rm p}$, we obtain
\begin{equation}
\label{eq:Cvnp}
c_{N} \approx c_{0} \left( \frac{n_{N}}{n_0} \right)^{1/3} T_9,
\end{equation}
with $N=$ n or p, and $c_{0} = 1.12\times
10^{20}$~erg~cm$^{-3}$~K$^{-1}$. In the case of very strong proton
superfluidity one has $c_{\rm p}\to 0$.

Integration of $Q$ and $c_{\text{core}}$ over the neutron star core
gives the neutrino luminosity $L_{\nu}$ and the heat capacity
$C_{\rm core}$ of the core. The same quantities for the crust are
negligible after the star reaches the state of internal thermal
relaxation (e.g. \citealt{Yak2001,YakPeth2004}). If so, $L_{\nu}$
and $C_{\rm core}$ are almost equal to the total neutrino luminosity
and heat capacity of the star, respectively. We will perform the
integration of $Q$ and $c_{\text{\rm{core}}}$ for different EOSs in
the core assuming isothermal interior of the star.

\section{Zoo of equations of state}
\newcommand{\nliiiwr}{NL3$\omega\rho$}

\begin{table}
\begin{center}
\caption{\label{tab:EOSparams}
The basic parameters of stars with the selected EOSs;
$M_{\mathrm{max}}$ and $R_{\mathrm{min}}$ refer to most massive stable stars;
$M_{\mathrm{DU}}$ and $R_{\mathrm{DU}}$ refer to stars where the DU process becomes allowed. }
\renewcommand{\arraystretch}{1.4}
\begin{tabular}{lcccccc}
\hline\hline
EOS  & $M_{\mathrm{max}}$, M$\odot$ & $R_{\mathrm{min}}$, km
& $M_{\mathrm{DU}}$, M$\odot$ & $R_{\mathrm{DU}}$, km \\
\hline
\nliiiwr & 2.75 & 13.00 & 2.60  & 13.79 \\
PAL4-240 & 1.93 & 10.24 & 1.64  & 11.93 \\
BSk21    & 2.27 & 11.01 & 1.59  & 12.59 \\
BSk20    & 2.16 & 10.10 &  ---  &  ---  \\
SLy      & 2.05 &  9.90 &  ---  &  ---  \\
APR II   & 1.92 & 10.20 & 1.89  & 10.83 \\
APR IV   & 2.16 & 10.82 & 1.73  & 12.48 \\
DDME2    & 2.48 & 12.05 &  ---  &  ---  \\
\hline
\end{tabular}
\end{center}
\end{table}

\begin{figure}
\includegraphics[width=\columnwidth]{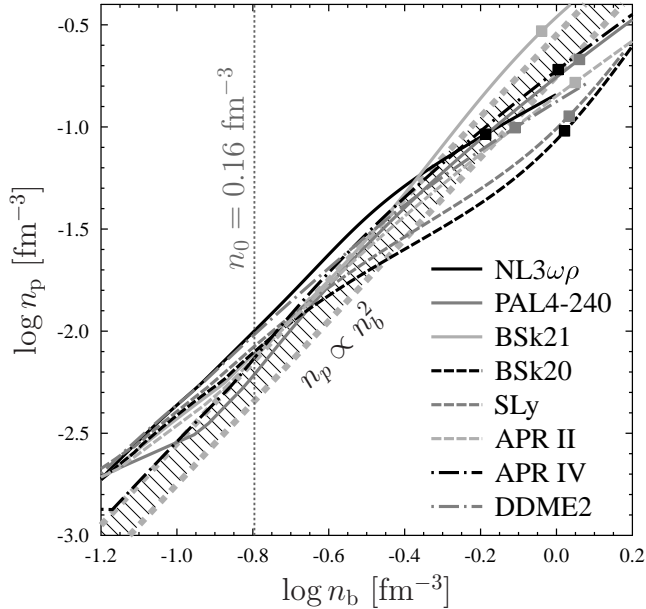}
\caption{\label{fig:np-nb} $n_{\rm p}-n_{\rm b}$ relations for the
selected EOSs. Squares mark the maximum $n_{\rm b}$ which is
possible in stable neutron stars with corresponding EOSs. The thick
shaded strip corresponds to the models which are similar
to the free-particle model. See text for details.}
\end{figure}

Let us take eight EOSs of superdense matter in neutron star cores.
They are illustrated in Figs.~\ref{fig:M-R}--\ref{fig:np-nb}. The
\nliiiwr\ and DDME2 EOSs are described in \cite{Fortin2016} and in
references therein;
the SLy EOS is taken from \cite{DH2001};
PAL4-240 is the model after \cite{PA1992} but with a different compression
modulus at saturation, $K_0 = 240$~MeV (see also the PAPAL model in
the Appendix~D of \citealt{HPY2007});
the APR II EOS is introduced by \citet{Gus2005};
the BSk20 and BSk21 EOSs are parametrized by \citet{BSk2013}, and
the APR IV EOS is constructed by \citet{KKPY14} (who called it the
HHJ EOS). For the SLy, BSk20 and BSk21 models, the EOSs in the crust
and the core are calculated in a unified way; the \nliiiwr\ and
DDME2 crustal parts are described by \cite{Fortin2016}; for other
models, the smooth composition EOS of the crust \citep{HPY2007} is
used. The most important parameters of neutron stars for the
selected EOSs are listed  in Table~\ref{tab:EOSparams}.

The $M(R)$ relations for neutron star models with these EOSs are
plotted in Fig.~\ref{fig:M-R}. We choose the EOSs with
different stiffness in order to consequently cover a large part of
the $M-R$ plane. Squares in Fig.~\ref{fig:M-R} correspond to the
most massive stable neutron star models. The selected EOSs are
reasonably consistent with recent discoveries of two massive ($M
\approx 2\,{\rm M}\odot$) neutron stars
\citep{Demorest2010,Antoniadis2013}. Circles mark configurations
where the DU process becomes allowed. Only five EOSs from
Table~\ref{tab:EOSparams} open the DU process before the most
massive stable configuration is reached.

In Fig.~\ref{fig:P-rho} we plot the $P-\rho$ relations for the
selected EOSs. These relations have several common features. First,
the EOSs at $\rho\sim \rho_0 = 2.8\times 10^{14}$~g~cm$^{-3}$
(the dotted vertical line) are not dramatically
different (differences in $P$ are within a factor of 2). It is
because, as a rule, the EOSs  are constructed in such a way to
reproduce the properties of saturated nuclear matter which are well
studied in laboratory. Secondly, the stiffer the high-density EOS,
the larger $M_{\rm max}$. Finally, in spite of the
similarity of the $P-\rho$ relations near $\rho_0$
they they are sufficiently different at $\rho\gtrsim2\rho_0$
which results in rather different $M-R$ relations. The straight
thick shaded strip line in Fig.~\ref{fig:P-rho} corresponds to
a family of simple polytropic EOS models
$P \propto \rho^{\gamma}$ with the power-law
index $\gamma=2.3$. It is a good overall approximation as discussed
in Section~\ref{sec:Integr-calibr}.

Fig.~\ref{fig:np-nb} illustrates another important property of the
selected EOSs, the relation between the proton $n_{\rm p}$ and total
baryon $n_{\rm b}$ number densities. A bunch of the curves for the
different EOSs is the thinnest at $n_{\rm b} \sim n_0$ (the
dotted vertical line). It is a consequence of the calibration
of the EOSs to the standard nuclear theory. The straight thick shaded strip corresponds to the relations $n_{\rm p} \propto n_{\rm b}^2$,
which can be derived from the free-particle model at not too high
$n_{\rm b}$ (e.g., \citealt{FM1979,ST1983}). Fig.~\ref{fig:np-nb}
shows that this simple approximation is qualitatively accurate which
is sufficient for our analysis.

Let us stress that we do not intend to accurately fit the EOSs
or number densities of different particles. Our aim is to suggest
some simple scaling expressions for these quantities and use
them to fit the expressions for the integral quantities, such as
$C_{\rm core}$ and $L_\nu^\infty$. One can treat these scaling
expressions as purely auxiliary and phenomenological (although we
prefer to introduce them on physical grounds). We will see that the
integration over the core absorbs the inaccuracy of scaling
expressions and enables us to accurately describe the integral
quantities.

\section{Integration of neutrino luminosity and heat capacity }
\newcommand{\Phiwave}{\widetilde{\Phi}}
\newcommand{\const}{{const}}
\newcommand{\Tg}{\widetilde{T}}
\newcommand{\Ts}{T_{\rm}}

\subsection{Basic expressions}
\label{sec:basicAppr}

The neutrino luminosity $L_{\nu}^\infty$ of a neutron star core
redshifted for a distant observer is given by
\begin{equation}
\label{eq:L-start}
L_{\nu}^\infty = \int_0^{R_{\rm core}} Q( \rho, T) \exp(2\Phi) \,
\frac{4\pi r^2 \,{\rm d}r}{\sqrt{1 - 2 G m/(r c^2)}}\,.
\end{equation}
The heat capacity of the core is
\begin{equation}
\label{eq:C-start}
C_{\rm core} = \int_0^{R_{\rm core}} c_{\text{core}}( \rho,T)\,
\frac{4\pi r^2\, {\rm d}r}{\sqrt{1 - 2 G m /(r c^2)}}.
\end{equation}
Here $R_{\rm core}$ is the core radius, $m = m(r)$ is the gravitational mass
inside the sphere of radial coordinate $r$, and $\Phi(r)$ is the metric function
determined by the equation (e.g. \citealt[Ch.~6]{HPY2007})
\begin{equation}
\label{eq:dPhi}
\frac{{\rm d}\Phi}{{\rm d}r} = -\frac{1}{ P + \rho c^2}\,\frac{{\rm d}P}{{\rm d}r}.
\end{equation}

The neutrino emissivity and the specific heat are expressed here as
functions of the local temperature $T$ and the local density $\rho$.
In a star with isothermal interiors $T(r)$ is given by equation
(\ref{eq:Tgrav}), $ T = \Tg \exp(-\Phi)$, with $\Tg$ being constant
over the isothermal region.

Let us analyse three cases of neutrino emission in equation
(\ref{eq:L-start}). The first is the SF case, where $Q = Q_{\rm nn}$
is given by equation (\ref{eq:Qnn}). The second is the non-SF case
with the forbidden DU process, so that $Q = Q_{\rm MU}$, equation
(\ref{eq:Qmu}). The third case is also for the non-SF core but with
the allowed DU process. In this case, we set $Q = Q_{\rm DU}$, given
by equation (\ref{eq:Qdu}). To simplify our analysis, in this paper
we use $Q = Q_{\rm DU}$ throughout the entire neutron star core (to
avoid complications associated with the introduction of the DU
threshold). This simplification is qualitatively justified because,
typically, $Q_{\rm DU}\sim 10^6 Q_{\rm MU}$, and even a small
central kernel with the allowed DU process makes $L_{\nu\,\rm
DU}^\infty$ drastically larger than $L_{\nu\,\rm MU}^\infty$.
However, it somewhat overestimates $L_{\nu\,\rm DU}^\infty$ and
gives only its firm upper limit.

As far as the specific heat is concerned, we consider two cases, the
non-SF and SF ones. They differ only by the presence or absence of
$c_{\rm p}$. Equations (\ref{eq:L-start}) and (\ref{eq:C-start})
allow us to numerically integrate $L_\nu^\infty$ and $C_{\rm
core}$ in the indicated cases for the selected EOSs.

\subsection{Analytic approximations of the integrals}
\label{sec:Integr-eval}

Exact analytic integration in equations (\ref{eq:L-start}) and
(\ref{eq:C-start}) is not possible. Instead, we derive approximate
expressions for these integrals and calibrate them using the results
of numerical integration.

Because the mass of a neutron star crust is typically about 1 per cent of
the total mass $M$, we can safely set that $m(R_{\rm core}) = M$ and
$\Phi(R_{\rm core}) = \Phi_{\rm surf}$, where
\begin{equation}
\label{eq:Phi_cc} \Phi_{\rm surf} = \ln \sqrt{1 - x_{\rm g}}, \quad
x_{\rm g}\equiv \frac{2G M}{R c^2}.
\end{equation}
It is convenient to introduce $\Phiwave = \Phi - \Phi_{\rm surf}$.

To proceed further we need approximate expressions for the number
densities of particles in neutron star cores. Let us assume that the
main contribution to the baryon number density $n_{\rm
b}=n_{\rm n}+n_{\rm p}$ is provided by neutrons, and the number
densities of charged particles are described by the model
similar to the free-particle one,
\begin{equation}
\label{eq:nnpe-nb}
  n_{\rm n} \approx n_{\rm b}, \quad n_{\rm p} \approx n_{\rm e}
  \approx a n_0 \left( \frac{n_{\rm b}}{n_0} \right)^2.
\end{equation}
Here $a$ is a dimensionless constant which can be treated as a value averaged
over the all selected EOSs. The $n_{\rm b} - \rho$ relation can be taken from \citet[Ch.~6]{HPY2007},
\begin{equation}
\label{eq:nb-rho}
n_{\rm b} = \frac{\rho}{m_{0}}\left( 1 + \frac{P}{\rho c^2} \right)\exp\Phiwave;
\end{equation}
$m_{0}$ being the rest mass per baryon in the $^{56}$Fe nucleus.

The approximate equality $n_{\rm e} \approx n_{\rm p}$ in equation
(\ref{eq:nnpe-nb}) can be significantly violated at very high
densities where $n_{\mu}\sim n_{\rm e}$. Such densities occur near
the centers of massive neutron stars; their contribution to the
integrated neutrino luminosities and heat capacities is small,
except for $L_{\nu\,\rm DU}$, where the contributions of the muon
and electron DU processes are equal.

As mentioned above, we study the three scenarios  (nn, MU and DU) of
neutrino emission in equation (\ref{eq:L-start}). In the first (nn)
case we take $Q=Q_{\rm nn}$ from equation (\ref{eq:Qnn}). In the
second (MU) case we employ $Q_{\rm MU}$ from equation
(\ref{eq:Qmu}), but we will additionally simplify it assuming
$\Omega(n_{\rm n,p,e,\mu}) = \const$. In the third (DU) case we use
equation (\ref{eq:Qdu}) but replace the sum of $\Theta$-functions by
a factor $2$.  Since typical densities, where the DU processes
operate, are so high that muons appear, such a simplification is
reasonable.

Considering the specific heat, we use the approximation
(\ref{eq:Cvnp}), $c_{\text{core}} = b c_{\rm n}$, with
different constants $b$ for the non-SF and SF cases.

Let us factorize (\ref{eq:L-start}) and (\ref{eq:C-start}) into
dimensional and dimensionless terms. It is convenient to define
\begin{subequations}
\label{eq:dimless}
\begin{equation}
\label{eq:dimless-x}
r = R_{\rm core} x, \qquad 0<x<1;
\end{equation}
\begin{equation}
\label{eq:dimless-f}
\rho(r) = \frac{M}{R_{\rm core}^3} f(x), \qquad f(0)\sim 1, \qquad f(1)\ll 1;
\end{equation}
\begin{equation}
\label{eq:dimless-F}
m(r) = M\, F(x), \qquad F(x) = 4\pi \int_0^x f(x'){x'}^2 \,{\rm d}x'.
\end{equation}
\end{subequations}
We assume that $f(x)$ has a universal form for any EOS, $M$ and $R$
of our study. According to  \citet{Latt2001}, such an approximation
is reasonable. Then equation (\ref{eq:L-start}) yields
\begin{multline}
\label{eq:L-dimlessInt}
 L_{\nu}^\infty = a'
\left( \frac{R_{\rm core}}{R} \right)^3
 Q_0 R^3 x_\rho^{k/3}
 \Tg_9^n \left( 1 - x_{\rm g} \right)^{1-n/2}  \\
 \times \int_0^1 x^2 \, {\rm d}x\,
 \left[ f(x)\left( 1 + \frac{P}{\rho c^2} \right) \right]^{k/3} \\
 \times \exp\left[ \left(\frac{k}{3}+2-n\right)\Phiwave\right]
 \left[ 1 - x_{\rm g}\,\frac{R}{R_{\rm core}}\frac{F(x)}{x}\right]^{-1/2},
\end{multline}
where $x_\rho=M/(\rho_0R^3)$; $k=1$, $n=8$ and $Q_0 = Q_{\rm nn\,0}$
for the  nn-bremsstrahlung; $k=2$, $n=8$ and $Q_0 = Q_{\rm MU\,0}$
for the MU process;  $k=2$, $n=6$ and $Q_0 = Q_{\rm DU\,0}$ for the
DU process; In equation (\ref{eq:L-dimlessInt}) we have introduced a
dimentionless constant $a'$ to absorb the inaccuracy of $L_{\nu}^{\infty}$
due to our approximations of $n_{\rm p}(n_{\rm b})$ and $\Omega$
in the DU and MU cases; in the SF case $a' = 1$. Similarly, for
the heat capacity (\ref{eq:C-start}) we obtain
\begin{multline}
\label{eq:C-dimlessInt}
    C_{\rm core} = b'
\left( \frac{R_{\rm core}}{R} \right)^3
 c_{0} R^3 x_\rho^{1/3} \Tg_9
        \left( 1 - x_{\rm g} \right)^{-1/2}  \\
    \times \int_0^1 x^2\,{\rm d}x\,
        \left[ f(x)\left( 1 + \frac{P}{\rho c^2} \right) \right]^{1/3}\\
        \times \exp\left(-{2 \over 3}\, \Phiwave \right)
    \left[ 1 - x_{\rm g}\,\frac{R}{R_{\rm core}}\frac{F(x)}{x}\right]^{-1/2},
\end{multline}
where constants $b'$ are different in the SF and non-SF cases.

Next consider a polytropic EOS, $P = P_0
\left(\rho/\rho_0\right)^{\gamma}$, with some effective $\gamma$
whose value will be obtained later by calibrating to numerical
calculations of $L_{\nu}^\infty$ and $C_{\rm core}$. Then we analytically
derive $\Phiwave$ from equation (\ref{eq:dPhi}) with the
condition $\Phiwave(R_{\rm core})=0$,
\begin{equation}
\label{eq:Phiwave}
 \Phiwave = -\frac{\gamma}{\gamma-1}
 \left[ \ln\left( 1 + \frac{P}{\rho c^2} \right) -
  \ln\left( 1 + \frac{P_{\rm cc}}{\rho_{\rm cc} c^2} \right)\right],
\end{equation}
where $\rho_{\rm cc} = \rho(R_{\rm core}) \approx \rho_0/2$
and $P_{\rm cc} = P(\rho_{\rm cc})$ are, respectively, the density and pressure at the core-crust interface. Actually, this solution behaves as
\begin{equation}
\label{eq:expPhiwave}
\exp{\Phiwave} \propto \left( 1 + \frac{P}{\rho c^2} \right)^{-{\gamma}/(\gamma-1)}.
\end{equation}

At the next step we stress that the ratio $R_{\rm core}/R = 0.8 -
1.0$ only slightly varies for different stellar masses higher than
$1\,{\rm M}\odot$. Thus we assume $R_{\rm core}/R\approx 0.9$ to be
constant in Eqs.~(\ref{eq:L-dimlessInt}) and
(\ref{eq:C-dimlessInt}). Then, combining equations
(\ref{eq:dimless-f}) and (\ref{eq:expPhiwave}) with
(\ref{eq:L-dimlessInt}) and (\ref{eq:C-dimlessInt}), we see that the
integrals are parametrised by $M$ and $R$. All uncertainties of
calculations of the integrals are encapsulated in the functions
$f(x)$ and $F(x)$. These functions should be smooth as proven by
(\ref{eq:dimless}). Thus the dependence of the integrals in
equations (\ref{eq:L-dimlessInt}) and (\ref{eq:C-dimlessInt}) on $M$
and $R$ can be understood using the midpoint method, by taking the
integrands at some fixed value of x between $0$ and $1$, $x_{\rm
mid}$. For simplicity, we assume that this value is independent of
$M$ and $R$. It is convenient to introduce
\begin{equation}
\label{eq:Jdef}
  J_{kp}(M,R) =  a_1 x_\rho^{k/3}
  \left( 1 - x_{\rm g} \right)^{-p/2}
    \frac{\left( 1 + a_3 x_\rho^{\gamma-1}
    \right)^{\frac{p\gamma - k/3}{\gamma-1}}}{\sqrt{1 - a_2 x_{\rm g} }}.
\end{equation}
Then the final expressions for the neutrino luminosity and heat
capacity take the forms
\begin{equation}
\label{eq:LC-final}
\left\{ \begin{array}{c}
L_{\nu}^{\infty} \\
C_{\rm core}
\end{array}\right\}
=  \left\{ \begin{array}{c}
Q_0 \\
c_{0}
\end{array}\right\}
R^3 \Tg_9^n J_{kp}(M,R).
\end{equation}
Here we use $R^3$ instead of $R_{\rm core}^3$ because $R_{\rm
core}/R$ is taken to be constant in our analytic models and the
difference is absorbed in fit parameters described below. The
exponents $n$, $p$ and $k$ are taken from equations
(\ref{eq:L-dimlessInt}) and (\ref{eq:C-dimlessInt}) and listed in
Table~\ref{tab:npkabcgamma}. The dimensionless parameters $a_1$,
$a_2$, $a_3$ and the most suitable values of $\gamma$ in
(\ref{eq:Jdef}) will be obtained by the calibration to numerical
calculations.

\subsection{Calibration to numerical calculations}
\label{sec:Integr-calibr}
\begin{table*}
\begin{center}
\caption{\label{tab:npkabcgamma}
Parameters of the approximations (\ref{eq:Jdef}) and (\ref{eq:LC-final})
for the neutrino luminosity $L_{\nu}^\infty$ and the heat capacity $C_{\rm core}$. For the luminosity, `nn' refers to the nn-bremsstrahlung (SF case),
while `MU' and `DU' correspond to the MU and DU processes (non-SF case),
respectively. For the heat capacity, the non-SF and SF cases differ by the
presence or absence of the proton contribution.  The two last columns give
root-mean-square (rms) relative errors and maximal relative errors
(indicating an EOS at which they occur, that always happens at $M=M_{\rm max}$).
See the text for details.}

\renewcommand{\arraystretch}{1.4}
\begin{tabular}{cccccccccccc}
\hline\hline
$L_\nu^\infty$ or $C_{\rm core}$ & Case & $Q_0$ or $c_{0}$ & $n$ & $k$ & $p$ & $a_1$ & $a_2$ & $a_3$ & $\gamma$ & rms & max error \\
\hline
          & nn (SF) & $1.77\times 10^{19}$ erg~cm$^{-3}$~s$^{-1}$ & 8 & 1 & 6 & 2.03 & 1.14 & 0.0047 &  2.51  & 0.14 & 0.54 at \nliiiwr  \\
$L_{\nu}^\infty$ & MU (non-SF) & $1.75\times 10^{21}$ erg~cm$^{-3}$~s$^{-1}$ & 8 & 2 & 6 & 1.12 & 1.14 & 0.0060 &  2.45  & 0.25 & 0.66 at \nliiiwr  \\
          & DU (non-SF) & $1.96\times 10^{27}$ erg~cm$^{-3}$~s$^{-1}$ & 6 & 2 & 4 & 1.01 & 1.14 & 0.0031 &  2.48  & 0.16 & 0.40 at \nliiiwr  \\
\hline
$C_{\rm core}$
& non-SF & $1.12\times 10^{20}$ erg~cm$^{-3}$~K$^{-1}$ & 1 & 1 & 1 & 2.68 & 1.14 & 0.0174 &  2.11  & 0.05 & 0.12 at BSk20 \\
          & SF & $1.12\times 10^{20}$ erg~cm$^{-3}$~K$^{-1}$ & 1 & 1 & 1 & 2.01 & 1.06 & 0.0159 &  2.18  & 0.05 & 0.09 at \nliiiwr \\
\hline
\end{tabular}
\end{center}
\end{table*}

\begin{figure*}
\includegraphics[width=0.495\textwidth]{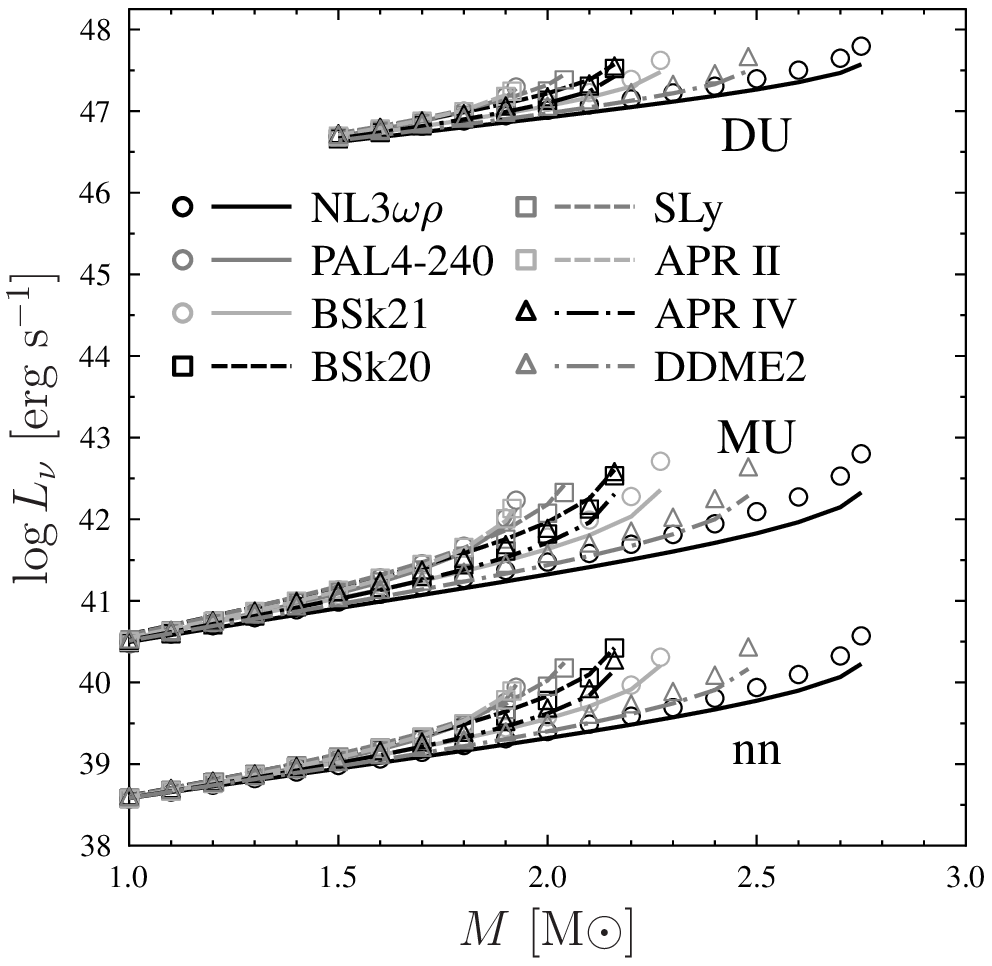}
\hfill
\includegraphics[width=0.495\textwidth]{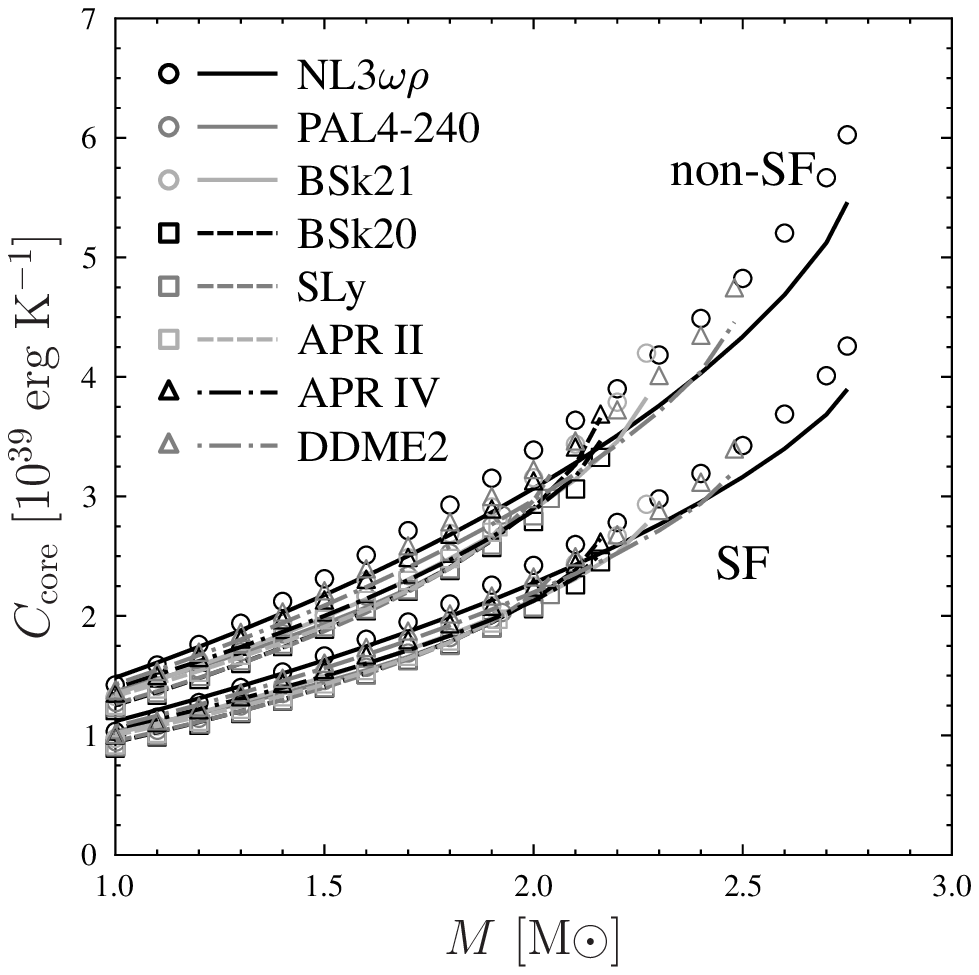}
\caption{\label{fig:M-LC} $L_\nu^\infty-M$ (left-hand panel) and
$C_{\rm core}-M$ (right-hand panel) relations for the seven selected
EOSs at $\Tg = 10^9$~K. Lines show the approximation
(\ref{eq:LC-final}); squares, circles and triangles present
numerical calculations. The labels `nn', `MU', `DU',  `SF' and
`non-SF' are the same as in Table~\ref{tab:npkabcgamma}. For $L_{\nu
\,\rm DU}^\infty$, the DU process is artificially extended over
the entire core, but the calculations are performed only at
$M\geqslant 1.5$~M$\odot$.}
\end{figure*}

The integrals (\ref{eq:L-start}) and (\ref{eq:C-start}) have been
calculated numerically. In this way we have obtained accurate
values of $L_\nu^\infty$ and $C_{\rm core}$ for any selected EOS,
any scenario (nn, MU, DU, SF, non-SF) and for a range of masses $M =
1.0\,{\rm M}{\odot},\, 1.1\,{\rm M}{\odot},\ldots M_{\rm max}$. In
the calculations, we have used the expressions for $Q$ and
$c_{\rm core}$ described in Sections~\ref{sec:2} and
\ref{sec:basicAppr}. As mentioned above, while calculating $L_{\nu
\rm DU}^\infty$ we have extended $Q_{\rm DU}$ over the entire
neutron star core. However, for the DU case we have not used
stellar models with $M<1.5$~M$\odot$ because $M_{\rm DU}> 1.5$~M$\odot$
for all our models (Table~\ref{tab:EOSparams}).

Our numerical results are shown by different symbols in
Fig.~\ref{fig:M-LC}. These data have been used to calibrate our
analytic approximation (\ref{eq:LC-final}). We have obtained 109
values of the neutrino luminosity (cases nn and MU, excluding DU) as
well as 109 values of the heat capacity (cases SF and non-SF). For
the DU case we have excluded 40 values with $M<1.5$~M$\odot$.
The trial fuctions $L_{\nu}^{\infty}(M,R)$ and $C_{\rm
core}(M,R)$ (equations (\ref{eq:Jdef}), (\ref{eq:LC-final}))
have been calibrated to these data sets. The target fuction to
minimize has been the relative root mean square (rms) error.
The optimised values of $a_1$, $a_2$, $a_3$ and $\gamma$ as
well as the corresponding fit errors are listed in
Table~\ref{tab:npkabcgamma}. The obtained approximations are
also plotted on Fig.~\ref{fig:M-LC}.

Let us discuss the approximations of $L_{\nu}^\infty$. They are most
precise for the nn-bremsstrahlung; the rms errors appear the
smallest because $Q_{\rm nn}$ is independent of the fractions of
charged particles in dense matter. However, it has large maximal
relative error which occurs for the most massive star with the
\nliiiwr\ EOS. This is because the approximation of this
EOS by a single polytrope does not reproduce well its high density
behaviour. The largest errors occur for the MU case due to a strong
dependence of $Q_{\rm MU}$ on the fractions of charged particles
through the factor $\Omega$. The approximation of $L_{\nu\,\rm
DU}^\infty$ is more accurate than the approximation of $L_{\nu\, \rm
MU}^\infty$ because $Q_{\rm DU}$ depends on $n_{\rm e}$ in a rather
simple way.

The importance of charged particle fractions can be
demonstrated by instructive examples of the BSk20 and APR~IV EOSs.
In Figs.~\ref{fig:M-R}--\ref{fig:M-LC} the corresponding curves are
plotted by short-dashed (BSk20) and dot-dashed (APR~IV) lines.
The initial numerical data in Fig.~\ref{fig:M-LC} are displayed
by black squares (BSk20) and triangles (APR~IV). According to
Fig.~\ref{fig:M-R}, these EOSs have very close maximum masses, but
the stars with the BSk20 EOS are more compact, i.e. have smaller
radii than the APR~IV stars of the same $M$. Roughly speaking, the
$M-R$ relations for these EOSs differ by a shift along the $R$ axis.
This means that a BSk20 star is denser than an APR~IV
star, and, therefore, has larger $L_\nu^\infty$. This is true for
$L_{\nu\, \rm nn}^\infty$ (Fig.~\ref{fig:M-LC}): black squares (for
BSk20) lie higher than black triangles (for APR~IV). This feature is
well reproduced by the black dashed and dot-dashed lines, which
show the approximation (\ref{eq:LC-final}) for these EOSs. In
contrast, the MU and DU luminosities are sensitive to the $n_{\rm p}
- n_{\rm b}$ relations. According to Fig.~\ref{fig:np-nb}, the
values of $n_{\rm p}$ for the APR~IV EOS are noticeably higher than
for the BSk20 EOS. The opposite effects of the two factors, the
greater compactness of the BSk20 stars and the larger $n_{\rm p}$
for the APR~IV stars, leads to their compensation. Accordingly, the
DU as well as the MU neutrino luminosities for these EOSs appear to
be close enough (triangles and black squares on the left-hand panel
of Fig.~\ref{fig:M-LC} overlap). Because the approximation
(\ref{eq:LC-final}) is derived using not very accurate description
of proton, electron and muon fractions, it cannot reproduce this
effect exactly; an approximate expression gives $L_{\nu\, \rm
DU}^\infty$ and $L_{\nu\, \rm MU}^\infty$ higher than numerical
values for the BSk20 EOS and lower than for the APR~IV EOS.

Another interesting note is that the parameter $a_1$ of our
approximation takes very close values for the non-SF (DU and MU)
cases but is several times larger for the SF (nn-bremsstrahlung)
case. This is because $a_1$ should include the value
$a^{1/3}$ from (\ref{eq:nnpe-nb}) and the extra factor
$f^{1/3}(x_{\rm mid})$ in the non-SF case but not in the SF case.
Because both these factors are smaller than 1, the value of $a_1$
should be several times lower in the non-SF case than in the SF one,
in agreement with Table~\ref{tab:npkabcgamma}.

Now let us outline the approximations of the heat capacity (the
right-hand side of Fig.~\ref{fig:M-LC}). The different EOSs give so
close values of $C_{\rm core}$, that the approximation
(\ref{eq:LC-final}) hardly resolves them.  Note that the parameters
$a_2$, $a_3$ and $\gamma$, which determine the $C_{\rm core}-M$
relation, are similar in the SF and non-SF cases. This supports the
idea that the behavior of the total specifiec heat is similar to
that of $c_{\rm n}$ in both these cases. A difference between the
values of $a_1$ shows that switching off the proton contribution by
superfluidity just reduces the heat capacity by about 25 per cent,
in agreement with the results by \citet{Page1993}.

Let us mention several common features of our approximations. First,
we can see that the index $\gamma$ ranges from about 2.1 to 2.5 with
the average value $\gamma \approx 2.3$. Such a polytropic EOS is
plotted in Fig.~\ref{fig:P-rho} by a thick green line and is in good
agreement with the selected EOSs. Secondly, the largest errors occur
at the maximum neutron star masses, because the higher the density
the stronger the difference between the EOS models. Thirdly, the
fact that the maximum errors occur for the \nliiiwr and BSk20
EOSs is explained by the largest deviations of $P-\rho$ (\nliiiwr)
and $n_{\rm p}-n_{\rm b}$ (BSk20) relations from the average trend
(Fig.~\ref{fig:np-nb}).

It is also remarkable that at low $M$ the exact $L_{\nu}^\infty-M$
dependence is insensitive to an EOS.  This gives hope to derive this
dependence analytically for $M\lesssim 1.5\,{\rm M}{\odot}$ which
is out the scope of the present work.

\section{Quiescent states of XRTs}
\label{sec:XRTs}

\newcommand{\Ldch}{L_{\mathrm{DCH}}}
\newcommand{\LdchZero}{L_{\mathrm{DCH}}^{(0)}}
\newcommand{\Mdot}{\langle\dot{M}\rangle}
\newcommand{\MdotTen}{\langle\dot{M}_{10}\rangle}

For illustration, we apply our approximations of $L_\nu^\infty$ to
analyse transiently accreting neutron stars in XRTs (low-mass X-ray
binaries). The formalism of applying the neutron star cooling theory
for exploring quasi-stationary thermal states of such objects in
quiescence is well known (e.g. \citealt{YLH2003,Lev2007}, also see
\citealt{YakPeth2004}). We will show that our approximations of
$L_\nu^\infty$ simplify an analysis of observations.

\subsection{Formalism of thermal states}

Let us outline the main features of thermal quasi-stationary
equilibrium of neutron stars in XRTs. During active states of such a
source, the neutron star accretes from its low-mass companion. The
accreted matter is compressed under the weight of newly accreted
material. It looks as if the accreted matter sinks into deeper
layers of the neutron star crust. This initiates nuclear
transformations of the accreted matter accompanied by the deep
crustal heating \citep{HZ1990,HZ2003,HZ2008} with the energy release
of about $1.5$~MeV per one accreted nucleon distributed mainly in
the inner crust. This deep crustal heating can be sufficiently
strong to warm up old transiently accreting stars and support their
observable thermal radiation during quiescent states of XRTs
\citep{Brown1998}.

We will not consider the episodes of rather long or intense
accretion when the deep heating is too intense and destroys
thermal equilibrium between the crust and the core
(e.g. \citealt{Nata2015} and references therein). We will restrict
our analysis to weaker or shorter accretion episodes. Then the
stellar interior remains isothermal and quasi-stationary.
On average, this heating is balanced by the thermal emission of photons from the
stellar surface and neutrinos from the stellar interior. In an
observer rest frame one has the thermal balance of the form
\begin{equation}
\label{eq:XRTbalance}
L_{\gamma}^{\infty} + L_{\nu}^{\infty} = L_{\mathrm{DCH}}^{\infty},
\end{equation}
where
\begin{equation}
\label{eq:LgammaInf}
L_{\gamma}^{\infty} = 4\pi\sigma R^2 T_{\rm s}^4 (1-x_{\rm g})
\end{equation}
is the redshifted photon luminosity of the star as a function of the
local effective surface temperature $T_{\rm s}$, $R$ and $M$
($x_{\rm g}$ is defined in equation (\ref{eq:Phi_cc})),
$L_{\nu}^{\infty}$ is the neutrino luminosity
(\ref{eq:L-start}) approximated by equation (\ref{eq:LC-final}),
and
\begin{equation}
\label{eq:LdchInf}
L_{\mathrm{DCH}}^{\infty} = \LdchZero \MdotTen \sqrt{1-x_{\rm g}}
\end{equation}
is the integrated rate of the energy release due to deep crustal
heating. The mass accretion rate $\MdotTen =
\Mdot/(10^{-10}\text{ M${\odot}$ yr$^{-1}$})$ has to be averaged
over the neutron star cooling time scales, and
$\LdchZero = 9.2\times 10^{33}$~erg~s$^{-1}$ is consistent with the
deep crustal heating rate (1.5 MeV per nucleon) given above.

Using equations (\ref{eq:XRTbalance})---(\ref{eq:LdchInf}) one can
calculate and plot the neutron star heating curves in the
$L_{\gamma}^{\infty}-\Mdot$ plane (e.g.,
\citealt{YakPeth2004,Lev2007}). Here we consider the two limiting
cases: (i) the photon cooling regime at $L_{\gamma}^\infty \gg
L_{\nu}^\infty$ and (ii) the neutrino cooling regime at
$L_{\nu}^\infty \gg L_{\gamma}^\infty$.

In the case (i) one has a simple $L_{\gamma}^{\infty}-\Mdot$ relation
\begin{equation}
\label{eq:Lgamma-Mdot_rad}
L_{\gamma}^{\infty} = \LdchZero \MdotTen \sqrt{1-x_g}.
\end{equation}
It slightly depends on $M$ and $R$ due to General Relativity effects.

In the case (ii) the relation between $L_\gamma$ and
$\langle\dot{M}\rangle$ is more complicated. One can use the
approximations (\ref{eq:LC-final}) to derive
$L_{\gamma}^{\infty}(\Mdot,M,R)$ analytically for the three
neutrino emission mechanisms, nn-bremsstrahlung in the SF case; MU or
DU processes in the non-SF case (Section~\ref{sec:Integr-eval}).
Then the neutron star thermal equilibrium reads
\begin{equation}
\label{eq:Ldch-Lnu} \LdchZero \MdotTen \sqrt{1-x_{\rm g}} = Q_0
R^3 J_{kp}(M,R) \Tg_9^n;
\end{equation}
$Q_0$, $n$, $k$ and $p$ are listed in Table~\ref{tab:npkabcgamma}
for each neutrino emission scenario. To calculate
$L_{\gamma}^{\infty}$ one needs to relate $T_{\rm s}$ and $\Tg$.
This can be done using the $T_{\rm s} - T_{\rm b}$ relations derived
by \cite{Pot2003}, where $T_{\rm b}$ is the temperature at the
bottom of the neutron star heat blanketing envelope. Making use of
$T_{\rm b}\sqrt{1-x_{\rm g}} = \Tg$ and equation
(\ref{eq:Ldch-Lnu}), we obtain
\begin{equation}
\label{eq:Tb-MdotMR}
T_{\rm b9} = \left( 1-x_{\rm g} \right)^{\frac{1-n}{2n}}
\left( \frac{\LdchZero \MdotTen}{Q_0 R^3 J_{kp}(M,R)} \right)^{1/n}.
\end{equation}

The $T_{\rm s} - T_{\rm b}$ relations are sensitive to the chemical
composition of the heat blanketing envelope which is rather
uncertain and depends on details of nuclear burning in the envelope.
The envelope can be almost purely iron if all the accreted material
has burnt to iron during an active XRT state. Alternatively, it can
be almost fully accreted or intermediate if the burning in the
envelope is slower. We consider the two limiting cases,
the case of non-accreted iron (Fe) envelope and the case of fully
accreted (acc) envelope. We denote the corresponding relations as
$T_{{\rm s}}(T_{\rm b}; M,R,j)$. Using equation
(\ref{eq:LgammaInf}) we have
\begin{equation}
\label{eq:LgammaInf-Mdot_final}
L_{\gamma}^{\infty} = 4\pi\sigma R^2 (1-x_{\rm g})
T_{\text{s}}^4 \left(T_{\rm b}; M,R,j \right),
\end{equation}
where $j =$ `acc' or `Fe' while $T_{\rm b}$ is given by equation (\ref{eq:Tb-MdotMR}).

\subsection{Model-independent analysis of thermal states}

\newcommand{\Sax}{SAX~J1808}
\newcommand{\HHH}{1H~1905}

\begin{table}
\begin{center}
\caption{\label{tab:XRTs} Accreting neutron stars in XRTs
whose
surface thermal emission in quasi-stationary quiescent states has
been detected/constrained
as plotted on the right-hand panel of Fig.\ \ref{fig:DoubleFigure}.}
\renewcommand{\arraystretch}{1.2}
\begin{tabular}{ll|ll}
\hline\hline
Num. & Source  &  Num.   & Source  \\
\hline
1    & 4U 1608--522     & 13    &  NGC 6440 X-1     \\
2    & Aql X-1          & 14    &  SAX J1810.8--2609 \\
3    & 4U 1730--22      & 15    &  MXB 1659--29  \\
4    & RX J1709--2639   & 16    &  IGR 00291+5934 \\
5    & Terzan 5         & 17    &  XTE J1814--338  \\
6    & 4U 2129+47       & 18    &  XTE J2123--058 \\
7    & 1M 1716--315     & 19    &  XTE J1807--294 \\
8    & Terzan 1         & 20    &  XTE J0929--314  \\
9    & 2S 1803--45      & 21    &  EXO 1747--214 \\
10   & KS 1731--260     & 22    &  NGC 6440 X-2  \\
11   & Cen X-4          & 23    &  1H 1905+000 \\
12   & XTE J1751--305   & 24    &  SAX J1808.4--3658 \\
\hline
\end{tabular}
\end{center}
\end{table}

\begin{figure*}
\includegraphics[width=\textwidth]{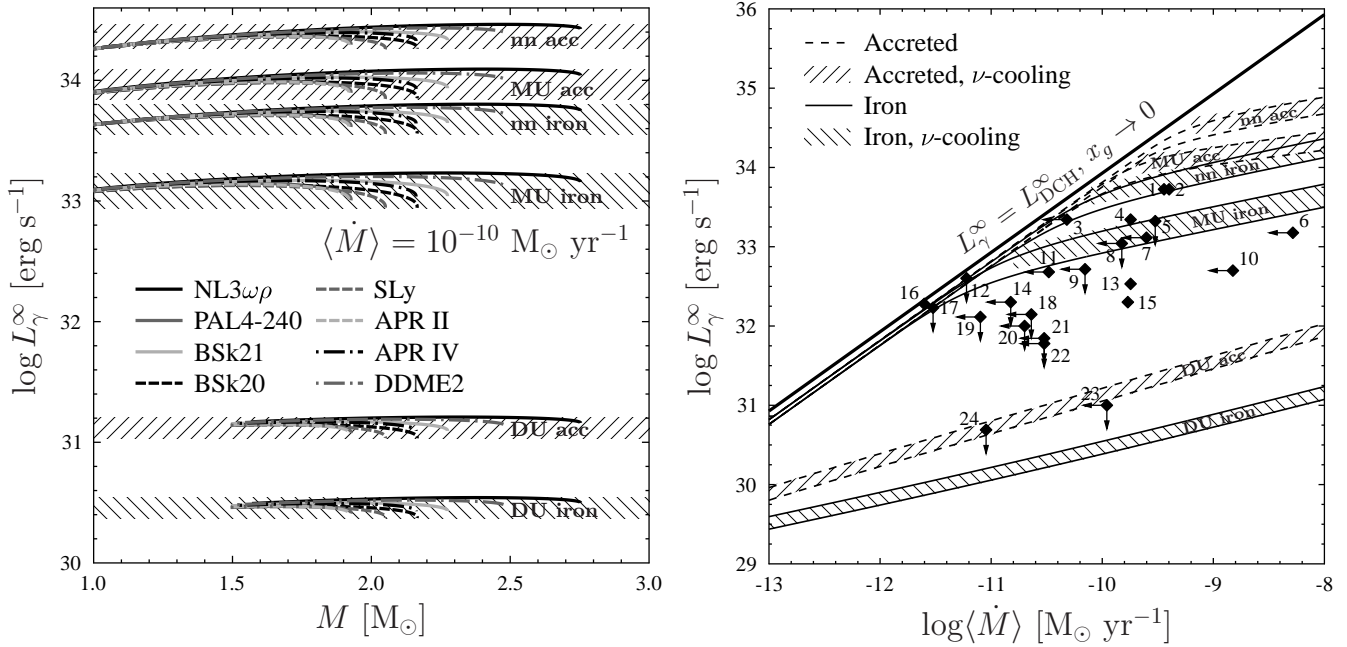}
\caption{\label{fig:DoubleFigure} Left-hand panel: Six families of
$L_{\gamma}^{\infty}-M$ relations for neutron stars in quiescent
states of XRTs with a fixed averaged mass accretion rate
$\Mdot=10^{-10}$~M$\odot$~yr$^{-1}$. The families are specified by
the neutrino emission mechanism (nn, MU or DU) and the composition of
the envelope (Fe or acc). The lines of different types are
calculated from equation (\ref{eq:LgammaInf-Mdot_final}) for the
different EOSs. The shaded bands enclose these lines. Right-hand
panel: 24 observed sources (Table~\ref{tab:XRTs}) in the
$L_\gamma^\infty-\Mdot$ plane. The bands are the same as on the
left-hand panel. Thin curves restrict numerical solutions of
equation~(\ref{eq:XRTbalance}) for the same six families. See
the text for details.}
\end{figure*}

The approximations (\ref{eq:LC-final}) and (\ref{eq:XRTbalance})
greatly simplify an analysis of thermal states of neutron stars in
XRTs. Now we need only mass $M$, radius $R$, the neutrino emission
scenario and the heat blanketing envelope type to calculate an
$L_{\gamma}^{\infty}-\Mdot$ relation.

The left-hand panel of Fig.~\ref{fig:DoubleFigure} shows six
families of $L_{\gamma}^{\infty}-M$ curves for a fixed time-averaged
mass accretion rate $\Mdot=10^{-10}$~M${\odot}$~yr$^{-1}$. The
families are for the three scenarios  of neutrino emission (nn, MU,
DU) and the two limiting models of neutron star heat blanketing
envelopes (Fe and acc). The curves of different types are calculated
from (\ref{eq:LgammaInf-Mdot_final}) for the different EOSs (Table
\ref{tab:EOSparams}) using the appropriate $M-R$ relations
(Fig.~\ref{fig:M-R}). The six horizontal shaded bands enclose the
curves of these six families. Recall that the DU curves are plotted
only at $M \geqslant 1.5$~M$\odot$, moreover, even at
$M \gtrsim 1.5$~M$\odot$ the DU curves are only underestimations
of $L_{\gamma}^{\infty}-\Mdot$ relations, as our approximations
overestimate $L_{\nu\,\rm DU}$. The presented bands can be treated as
overestimated photon thermal luminosity bands for the DU curves.
For each band, the bounding stellar models are chosen as
$M = 2.05$~M${\odot}$, $R = 9.90$~km (SLy; lower bound) and
$M = 2.4$~M${\odot}$, $R = 14.0$~km (\nliiiwr; upper bound).

Note that the $L_{\gamma}^{\infty}-M$ relations are
mostly non-monotonic. The luminosity increases with growing $M$ for
low-mass stars but then decreases again for massive stars; the
lowest $M$ does not necessarily correspond to the brightest source in the
band. This is a consequence of General Relativity effects.

The right-hand panel of Fig.~\ref{fig:DoubleFigure} shows the same
six bands as on the left-hand panel, but in the
$L_{\gamma}^{\infty}-\Mdot$ plane. Thin curves present numerical
solutions of the initial equation (\ref{eq:XRTbalance})
(without assuming $L_{\nu}^\infty \gg
L_{\gamma}^\infty $) for the bounding neutron star models (see
above). Such solutions almost coincide with the approximate ones
if $L_{\gamma}^{\infty} \lesssim 0.1 \Ldch^\infty$.
We stress that Fig.\ \ref{fig:DoubleFigure} illustrates only six
representative scenarios of thermal states of neutron stars. Any
filled band limits possible values of $L_\gamma^\infty$ versus $M$ or
$\Mdot$ for neutron stars with the selected EOSs. Although the
set of these EOSs is restricted we expect that the bands are robust
(should not be greatly changed for a wider class of EOSs of nucleon
matter). Another important feature is that the direct effect of the
EOS on thermal states of neutron stars in XRTs for any of the six
scenarios does not seem very strong. Therefore, the analysis of these
scenarios is relatively model-independent.

Note also that the thin theoretical curves are valid only at $\Mdot
\lesssim 10^{-9}$~M$\odot$~yr$^{-1}$. At higher accretion rates the
internal thermal equilibrium of the neutron star is violated. This
leads to higher $L_{\gamma}^{\infty}$ than predicted by isothermal
calculations.

The theoretical bands on the right-hand panel of Fig.\
\ref{fig:DoubleFigure} are compared with observational data (or
observational upper limits shown by arrows) for 24 sources. The data
are the same as those presented by \citet{Bez2015,Bez2015a} (who
gave also the list of original publications). The sources are
numbered in accordance with Table \ref{tab:XRTs}. They are mainly
located between the MU and DU-cooling bands. Note the absence of any source with measured $\Mdot > 10^{-9}$~M$\odot$~yr$^{-1}$ except for 4U~2129+47 (source 6) and KS~1731--260 (source 10), for which only
$\Mdot$ upper limits are given.

The thick solid black line on the right-hand panel of Fig.~\ref{fig:DoubleFigure} corresponds to $L_\gamma=L_{\rm DCH}$ (for non-redshifted quantities, $x_{\rm g}\to 0$). It is the absolute upper bound on the thermal quiescent luminosity of a neutron star in the deep crustal heating scenario. One can see that all the data satisfy this upper bound. Actually, the
data agree also with the highest `nn acc' band and even with lower `MU acc' or `nn iron' bands, but disagree with the `MU iron' band. This means that all the hottest neutron stars observed in quescent states of XRTs cannot be explained by the standard MU neutrino cooling and the standard iron envelopes.
One needs either a slower (nn) neutrino cooling (strong SF) and/or accreted envelopes to raise $L_\gamma^\infty$ and explain the data. The possibility of raising $L_\gamma^\infty$ by strong SF has been analysed earlier (e.g., \citealt{YakPeth2004}).

While the direct effects of EOS on thermal states of neutron stars
in XRTs are not strong, other factors are seen
(Fig.\ \ref{fig:DoubleFigure}) to affect these
thermal states much stronger. For instance, fixing
the neutrino emission scenario (nn, MU or DU) but varying chemical
composition of the heat blanketing envelope from pure iron to
pure accreted  can produce much stronger variations of
$L_\gamma^\infty$ (much wider bands) than those due to the EOSs.
Alternatively, fixing the envelope composition (Fe or acc) and
varying the neutrino emission scenarios (from nn to MU by proton
superfluidity and to DU either by superfluidity or by nuclear
physics effects, which shift $M_{\rm DU}$), one can produce even
stronger variations of $L_\gamma^\infty$.

Recall that in the strong neutrino emission (DU) scenario we have
artificially extended the operation of the DU process over the entire neutron star core overestimating $L_{\nu\,\rm DU}^\infty$ in massive stars. Accordingly, our DU bands in Fig.\ \ref{fig:DoubleFigure} appear to be downshfted by a factor of few with respect to the heating curves calculated in the previous analyses (e.g., \citealt{YLH2003,Lev2007,Bez2015,Bez2015a}). Therefore, one should be careful in using our DU bands for analysing thermal states of the coldest sources, 1H~1905+000 \citep{Jonker06,JSCJ07b,Heinke09b} and SAX~J1808.4--3658 \citep{Heinke07,TGK05,JWK04} as massive neutron stars with the DU process on. These sources are very important for proving (or disproving) the operation of the DU process in massive stars. The real DU bands should lie by a factor of $\sim 3$ higher (in natural scale). However, we believe that our current DU bands have realistic widths and reasonably well describe variations of $L_\gamma^\infty$ due to composition of the envelopes. We expect to improve our model-independent analysis of the DU bands in our future publication.

\section{Discussion and conclusions}

We have considered a representative set of seven EOSs for neutron stars
with nucleon cores (Table~\ref{tab:EOSparams}) and analysed the
models of neutron stars with isothermal interiors (with constant
redshifted internal temperatures $\widetilde{T}$). We have
calculated the neutrino luminosities $L_\nu^\infty$ and heat
capacities $C_{\rm core}$ of the cores of these stars with masses
$M=1.1\,{\rm M}\odot,~1.2\,{\rm M}\odot,\ldots,M_{\rm max}$ for
several important scenarios of neutron star internal structure. The
quantities $L_\nu^\infty$ and $C_{\rm core}$ almost coincide with
the total neutrino luminosities and heat capacities of neutron
stars.

Specifically, we have studied the three scenarios for $L_\nu^\infty$
which correspond to (i) the neutrino-pair bremsstrahlung in nn
collisions (owing to the presence of strong proton superfluidity in
the core); (ii) non-SF stars which cool via MU processes; (iii)
non-SF stars cooling via powerful DU processes. We have considered
the two scenarios for $C_{\rm core}$, relevant for non-SF cores and
the cores with strong proton superfluidity. The calculated values of
$L_\nu^\infty$ and $C_{\rm core}$ have been accurately fitted by the
analytic expressions (\ref{eq:LC-final}) which are universal for all
selected EOSs. The fit parameters (Table~\ref{tab:npkabcgamma})
are almost independent of the specific EOS. We expect that
$L_\nu^\infty$ and $C_{\rm core}$ calculated for neutron stars with
other EOSs of nucleon matter would be similar and could be
approximated in the same way making our approximations almost model
independent.

In this sense our consideration extends model-independent analysis
of cooling neutron stars based on the standard neutrino cooling
function $\ell(\widetilde{T})=L_{\nu \, \rm MU}^\infty
(\widetilde{T})/C_{\rm core}^{\rm non-SF}(\widetilde{T})$
\citep{Yak2011,Weisskopf_etal11,Kloch2015,Of2015-NS}.
\citet{Of2015-NS} derived also model-independent approximations to
the neutrino cooling function $\ell(\widetilde{T})=L_{\nu\, \rm
nn}^\infty (\widetilde{T})/C_{\rm core}^{\rm SF}(\widetilde{T})$ for
stars with strong proton superfluidity. \citet{SY2015} performed a
more complicated model-independent analysis of the cooling enhanced
by the onset of triplet-state pairing of neutrons and associated
neutrino emission in neutron star cores.

Our present results are more refined because we analyse
a weak dependence of $L_\nu^\infty$ and $C_{\rm core}$ on the EOS.
Our approximations of $L_\nu^\infty$ and $C_{\rm core}$ greatly
simplify calculations of cooling of isolated neutron stars at the
neutrino cooling stage after the initial thermal relaxation ($100
\lesssim t \lesssim 10^5$ yr). This cooling is governed by
the neutrino cooling function $\ell=L_{\nu}^\infty / C_{\rm core}$.
The expressions for $\ell$ obtained from our fits of
$L_\nu^\infty$ and $C_{\rm core}$ are in good agreement with the
approximations used in \citet{Yak2011,Weisskopf_etal11,Kloch2015,Of2015-NS}
but ours seem more complete. The approximations of
$C_{\rm core}$ should simplify cooling simulations of isolated neutron stars at
the photon cooling stage (at $t\gtrsim 10^5$ yr, when the neutrino luminosity
$L_\nu^\infty$ becomes unimportant). Finally, the approximations we obtained for
$L_\nu^\infty$ simplify considerations of thermal states of accreting
neutron stars in quasi-stationary XRTs as demonstrated in Section \ref{sec:XRTs}.

Let us stress that our results are far from being perfect because they
are obtained under a number of simplified assumptions.
For instance, while constructing the approximations we have assumed
the ratio $R_{\rm core}/R\approx 0.9$ to be constant.
In fact, it slightly decreases with the growth of $M$.
It may be so that one can improve our fits taking into account
the approximations of {$R_{\rm core}/R$} by \citet{ZH2016}.

While approximating $L_{\nu\, \rm DU}^\infty$  we have assumed that the
DU process is open in the entire core. In fact,
it can operate in a small central kernel. The size of this
kernel depends on $M$ and the EOS model. Its radius is
zero at $M = M_{\mathrm{DU}}$ but increases with growing $M$.
Moreover, according to Table~\ref{tab:EOSparams}, for three of the seven selected EOSs the DU process is forbidden in all stable neutron star models.
Thus, our approximation of $L_{\nu\, \rm DU}^\infty$ can be treated
as an overestimation to be improved in the future.

We have considered superfluidity of nucleons in a simplified
manner, i.e. we have assumed that neutrons are totally non-superfluid but
protons are either totally non-superfluid or fully superfluid.
The advantages and disadvantages of such a treatment are discussed by
\citet{Of2015-NS}.

Another simplification of our approach is in using constant effective nucleon
masses equal to 0.7 of their bare masses. In addition, the neutrino
emissivities have been calculated employing approximate squared matrix
elements of neutrino reactions taken by \cite{Yak2001} from calculations by
\cite{FM1979}. We expect that we can include a more advanced physics using a
similar formalism when this physics appears for a number of representative EOSs.

\section*{Acknowlegements}

The work of DG was supported partly by the RFBR (grants
14-02-00868-a and 16-29-13009-ofi-m) and the work of PH, LZ and MF
by the Polish NCN research grant no. 2013/11/B/ST9/04528. One of the
authors (DDO) is grateful to N. Copernicus Astronomical Center for
hospitality and perfect working conditions.


\begin{thebibliography}{}

\bibitem[\protect\citeauthoryear{{Antoniadis}, {Freire}, {Wex}, {Tauris},
  {Lynch}, {van Kerkwijk} \& {et al.}}{{Antoniadis}
  et~al.}{2013}]{Antoniadis2013}
{Antoniadis} J.,  {Freire} P.~C.~C.,  {Wex} N.,  {Tauris} T.~M.,  {Lynch}
  R.~S.,  {van Kerkwijk} M.~H.,    {et al.} 2013, Science, 340, 448

\bibitem[\protect\citeauthoryear{{Beznogov} \& {Yakovlev}}{{Beznogov} \&
  {Yakovlev}}{2015a}]{Bez2015}
{Beznogov} M.~V.,  {Yakovlev} D.~G.,  2015a, MNRAS, 447, 1598

\bibitem[\protect\citeauthoryear{{Beznogov} \& {Yakovlev}}{{Beznogov} \&
  {Yakovlev}}{2015b}]{Bez2015a}
{Beznogov} M.~V.,  {Yakovlev} D.~G.,  2015b, MNRAS, 452, 540

\bibitem[\protect\citeauthoryear{{Brown}, {Bildsten} \& {Rutledge}}{{Brown}
  et~al.}{1998}]{Brown1998}
{Brown} E.~F.,  {Bildsten} L.,    {Rutledge} R.~E.,  1998, Atrophys. J. Lett.,
  504, L95

\bibitem[\protect\citeauthoryear{{Degenaar}, {Wijnands}, {Bahramian},
  {Sivakoff}, {Heinke}, {Brown}, {Fridriksson}, {Homan}, {Cackett}, {Cumming},
  {Miller}, {Altamirano} \& {Pooley}}{{Degenaar} et~al.}{2015}]{Nata2015}
{Degenaar} N.,  {Wijnands} R.,  {Bahramian} A.,  {Sivakoff} G.~R.,  {Heinke}
  C.~O.,  {Brown} E.~F.,  {Fridriksson} J.~K.,  {Homan} J.,  {Cackett} E.~M.,
  {Cumming} A.,  {Miller} J.~M.,  {Altamirano} D.,    {Pooley} D.,  2015,
  MNRAS, 451, 2071

\bibitem[\protect\citeauthoryear{{Demorest}, {Pennucci}, {Ransom}, {Roberts} \&
  {Hessels}}{{Demorest} et~al.}{2010}]{Demorest2010}
{Demorest} P.~B.,  {Pennucci} T.,  {Ransom} S.~M.,  {Roberts} M.~S.~E.,
  {Hessels} J.~W.~T.,  2010, Nature, 467, 1081

\bibitem[\protect\citeauthoryear{{Douchin} \& {Haensel}}{{Douchin} \&
  {Haensel}}{2001}]{DH2001}
{Douchin} F.,  {Haensel} P.,  2001, Astron. Astrophys., 380, 151

\bibitem[\protect\citeauthoryear{{Fortin}, {Providencia}, {Raduta},
  {Gulminelli}, {Zdunik}, {Haensel} \& {Bejger}}{{Fortin}
  et~al.}{2016}]{Fortin2016}
{Fortin} M.,  {Providencia} C.,  {Raduta} A.~R.,  {Gulminelli} F.,  {Zdunik}
  J.~L.,  {Haensel} P.,    {Bejger} M.,  2016, ArXiv e-prints

\bibitem[\protect\citeauthoryear{{Friman} \& {Maxwell}}{{Friman} \&
  {Maxwell}}{1979}]{FM1979}
{Friman} B.~L.,  {Maxwell} O.~V.,  1979, Astrophys. J., 232, 541

\bibitem[\protect\citeauthoryear{{Glen} \& {Sutherland}}{{Glen} \&
  {Sutherland}}{1980}]{GS1980}
{Glen} G.,  {Sutherland} P.,  1980, Astrophys. J., 239, 671

\bibitem[\protect\citeauthoryear{{Gudmundsson}, {Pethick} \&
  {Epstein}}{{Gudmundsson} et~al.}{1983}]{GPE1983}
{Gudmundsson} E.~H.,  {Pethick} C.~J.,    {Epstein} R.~I.,  1983, Astrophys.
  J., 272, 286

\bibitem[\protect\citeauthoryear{{Gusakov}, {Kaminker}, {Yakovlev} \&
  {Gnedin}}{{Gusakov} et~al.}{2005}]{Gus2005}
{Gusakov} M.~E.,  {Kaminker} A.~D.,  {Yakovlev} D.~G.,    {Gnedin} O.~Y.,
  2005, MNRAS, 363, 555

\bibitem[\protect\citeauthoryear{{Haensel}, {Potekhin} \& {Yakovlev}}{{Haensel}
  et~al.}{2007}]{HPY2007}
{Haensel} P.,  {Potekhin} A.~Y.,    {Yakovlev} D.~G.,  2007, {Neutron Stars. 1.
  Equation of State and Structure}.
Springer, New York

\bibitem[\protect\citeauthoryear{{Haensel} \& {Zdunik}}{{Haensel} \&
  {Zdunik}}{1990}]{HZ1990}
{Haensel} P.,  {Zdunik} J.~L.,  1990, Astron. Astrophys., 227, 431

\bibitem[\protect\citeauthoryear{{Haensel} \& {Zdunik}}{{Haensel} \&
  {Zdunik}}{2003}]{HZ2003}
{Haensel} P.,  {Zdunik} J.~L.,  2003, Astron. Astrophys., 404, L33

\bibitem[\protect\citeauthoryear{{Haensel} \& {Zdunik}}{{Haensel} \&
  {Zdunik}}{2008}]{HZ2008}
{Haensel} P.,  {Zdunik} J.~L.,  2008, Astron. Astrophys., 480, 459

\bibitem[\protect\citeauthoryear{{Heinke}, {Jonker}, {Wijnands}, {Deloye} \&
  {Taam}}{{Heinke} et~al.}{2009}]{Heinke09b}
{Heinke} C.~O.,  {Jonker} P.~G.,  {Wijnands} R.,  {Deloye} C.~J.,    {Taam}
  R.~E.,  2009, Astrophys. J., 691, 1035

\bibitem[\protect\citeauthoryear{{Heinke}, {Jonker}, {Wijnands} \&
  {Taam}}{{Heinke} et~al.}{2007}]{Heinke07}
{Heinke} C.~O.,  {Jonker} P.~G.,  {Wijnands} R.,    {Taam} R.~E.,  2007,
  Astrophys. J., 660, 1424

\bibitem[\protect\citeauthoryear{{Jonker}, {Bassa}, {Nelemans}, {Juett},
  {Brown} \& {Chakrabarty}}{{Jonker} et~al.}{2006}]{Jonker06}
{Jonker} P.~G.,  {Bassa} C.~G.,  {Nelemans} G.,  {Juett} A.~M.,  {Brown} E.~F.,
     {Chakrabarty} D.,  2006, MNRAS, 368, 1803

\bibitem[\protect\citeauthoryear{{Jonker}, {Steeghs}, {Chakrabarty} \&
  {Juett}}{{Jonker} et~al.}{2007}]{JSCJ07b}
{Jonker} P.~G.,  {Steeghs} D.,  {Chakrabarty} D.,    {Juett} A.~M.,  2007,
  Astrophys. J. Lett., 665, L147

\bibitem[\protect\citeauthoryear{{Jonker}, {Wijnands} \& {van der
  Klis}}{{Jonker} et~al.}{2004}]{JWK04}
{Jonker} P.~G.,  {Wijnands} R.,    {van der Klis} M.,  2004, MNRAS, 349, 94

\bibitem[\protect\citeauthoryear{{Kaminker}, {Kaurov}, {Potekhin} \&
  {Yakovlev}}{{Kaminker} et~al.}{2014}]{KKPY14}
{Kaminker} A.~D.,  {Kaurov} A.~A.,  {Potekhin} A.~Y.,    {Yakovlev} D.~G.,
  2014, MNRAS, 442, 3484

\bibitem[\protect\citeauthoryear{{Kaminker}, {Yakovlev} \&
  {Haensel}}{{Kaminker} et~al.}{2016}]{KYH2016}
{Kaminker} A.~D.,  {Yakovlev} D.~G.,    {Haensel} P.,  2016, Astrophys. Sp.
  Sci., 361, 267

\bibitem[\protect\citeauthoryear{{Klochkov}, {Suleimanov}, {P{\"u}hlhofer},
  {Yakovlev}, {Santangelo} \& {Werner}}{{Klochkov} et~al.}{2015}]{Kloch2015}
{Klochkov} D.,  {Suleimanov} V.,  {P{\"u}hlhofer} G.,  {Yakovlev} D.~G.,
  {Santangelo} A.,    {Werner} K.,  2015, Astron. Astrophys., 573, A53

\bibitem[\protect\citeauthoryear{{Lattimer}, {Pethick}, {Prakash} \&
  {Haensel}}{{Lattimer} et~al.}{1991}]{LPPH1991}
{Lattimer} J.~M.,  {Pethick} C.~J.,  {Prakash} M.,    {Haensel} P.,  1991,
  Physical Review Letters, 66, 2701

\bibitem[\protect\citeauthoryear{{Lattimer} \& {Prakash}}{{Lattimer} \&
  {Prakash}}{2001}]{Latt2001}
{Lattimer} J.~M.,  {Prakash} M.,  2001, Astrophys. J, 550, 426

\bibitem[\protect\citeauthoryear{{Levenfish} \& {Haensel}}{{Levenfish} \&
  {Haensel}}{2007}]{Lev2007}
{Levenfish} K.~P.,  {Haensel} P.,  2007, Astrophys. Sp. Sci., 308, 457

\bibitem[\protect\citeauthoryear{{Ofengeim}, {Kaminker}, {Klochkov},
  {Suleimanov} \& {Yakovlev}}{{Ofengeim} et~al.}{2015}]{Of2015-NS}
{Ofengeim} D.~D.,  {Kaminker} A.~D.,  {Klochkov} D.,  {Suleimanov} V.,
  {Yakovlev} D.~G.,  2015, MNRAS, 454, 2668

\bibitem[\protect\citeauthoryear{{Page}}{{Page}}{1993}]{Page1993}
{Page} D.,  1993, in {Guidry} M.~W.,  {Strayer} M.~R.,  eds, Nuclear Physics in
  the Universe {The Geminga neutron star: Evidence for nucleon superfluidity at
  very high density}.
pp 151--162

\bibitem[\protect\citeauthoryear{{Page} \& {Applegate}}{{Page} \&
  {Applegate}}{1992}]{PA1992}
{Page} D.,  {Applegate} J.~H.,  1992, Astrophys. J. Lett., 394, L17

\bibitem[\protect\citeauthoryear{{Page}, {Lattimer}, {Prakash} \&
  {Steiner}}{{Page} et~al.}{2009}]{PLPS2009}
{Page} D.,  {Lattimer} J.~M.,  {Prakash} M.,    {Steiner} A.~W.,  2009,
  Astrophys. J., 707, 1131

\bibitem[\protect\citeauthoryear{{Potekhin}, {Chabrier} \&
  {Yakovlev}}{{Potekhin} et~al.}{1997}]{PCY1997}
{Potekhin} A.~Y.,  {Chabrier} G.,    {Yakovlev} D.~G.,  1997, Astron.
  Astrophys., 323, 415

\bibitem[\protect\citeauthoryear{{Potekhin}, {Fantina}, {Chamel}, {Pearson} \&
  {Goriely}}{{Potekhin} et~al.}{2013}]{BSk2013}
{Potekhin} A.~Y.,  {Fantina} A.~F.,  {Chamel} N.,  {Pearson} J.~M.,
  {Goriely} S.,  2013, Astron. Astrophys., 560, A48

\bibitem[\protect\citeauthoryear{{Potekhin}, {Pons} \& {Page}}{{Potekhin}
  et~al.}{2015}]{Pot2015}
{Potekhin} A.~Y.,  {Pons} J.~A.,    {Page} D.,  2015, Space Sci. Rev., 191, 239

\bibitem[\protect\citeauthoryear{{Potekhin}, {Yakovlev}, {Chabrier} \&
  {Gnedin}}{{Potekhin} et~al.}{2003}]{Pot2003}
{Potekhin} A.~Y.,  {Yakovlev} D.~G.,  {Chabrier} G.,    {Gnedin} O.~Y.,  2003,
  Astrophys. J., 594, 404

\bibitem[\protect\citeauthoryear{{Shapiro} \& {Teukolsky}}{{Shapiro} \&
  {Teukolsky}}{1983}]{ST1983}
{Shapiro} S.~L.,  {Teukolsky} S.~A.,  1983, {Black holes, white dwarfs, and
  neutron stars: The physics of compact objects}.
Wiley-Interscience, New York

\bibitem[\protect\citeauthoryear{{Shternin} \& {Yakovlev}}{{Shternin} \&
  {Yakovlev}}{2015}]{SY2015}
{Shternin} P.~S.,  {Yakovlev} D.~G.,  2015, MNRAS, 446, 3621

\bibitem[\protect\citeauthoryear{{Tomsick}, {Gelino} \& {Kaaret}}{{Tomsick}
  et~al.}{2005}]{TGK05}
{Tomsick} J.~A.,  {Gelino} D.~M.,    {Kaaret} P.,  2005, Astrophys. J., 635,
  1233

\bibitem[\protect\citeauthoryear{{Weisskopf}, {Tennant}, {Yakovlev}, {Harding},
  {Zavlin}, {O'Dell}, {Elsner} \& {Becker}}{{Weisskopf}
  et~al.}{2011}]{Weisskopf_etal11}
{Weisskopf} M.~C.,  {Tennant} A.~F.,  {Yakovlev} D.~G.,  {Harding} A.,
  {Zavlin} V.~E.,  {O'Dell} S.~L.,  {Elsner} R.~F.,    {Becker} W.,  2011,
  Astrophys. J., 743, 139

\bibitem[\protect\citeauthoryear{{Yakovlev} \& {Haensel}}{{Yakovlev} \&
  {Haensel}}{2003}]{YakHaens2003}
{Yakovlev} D.~G.,  {Haensel} P.,  2003, Astron. Astrophys., 407, 259

\bibitem[\protect\citeauthoryear{{Yakovlev}, {Ho}, {Shternin}, {Heinke} \&
  {Potekhin}}{{Yakovlev} et~al.}{2011}]{Yak2011}
{Yakovlev} D.~G.,  {Ho} W.~C.~G.,  {Shternin} P.~S.,  {Heinke} C.~O.,
  {Potekhin} A.~Y.,  2011, MNRAS, 411, 1977

\bibitem[\protect\citeauthoryear{{Yakovlev}, {Kaminker}, {Gnedin} \&
  {Haensel}}{{Yakovlev} et~al.}{2001}]{Yak2001}
{Yakovlev} D.~G.,  {Kaminker} A.~D.,  {Gnedin} O.~Y.,    {Haensel} P.,  2001,
  Phys. Rep., 354, 1

\bibitem[\protect\citeauthoryear{{Yakovlev}, {Levenfish} \&
  {Haensel}}{{Yakovlev} et~al.}{2003}]{YLH2003}
{Yakovlev} D.~G.,  {Levenfish} K.~P.,    {Haensel} P.,  2003, Astron.
  Astrophys., 407, 265

\bibitem[\protect\citeauthoryear{{Yakovlev} \& {Pethick}}{{Yakovlev} \&
  {Pethick}}{2004}]{YakPeth2004}
{Yakovlev} D.~G.,  {Pethick} C.~J.,  2004, Annu. Rev. Astron. Astrophys., 42,
  169

\bibitem[\protect\citeauthoryear{{Zdunik}, {Fortin} \& {Haensel}}{{Zdunik}
  et~al.}{2016}]{ZH2016}
{Zdunik} J.~L.,  {Fortin} M.,    {Haensel} P.,  2016, {\it in prep}., Astron.
  Astrophys.

\end{thebibliography}
\end{document}